\documentclass[pra,preprint,superscriptaddress,amssymb]{revtex4}
\usepackage{graphicx}
\usepackage{bm}
\usepackage{amsmath,amssymb,latexsym}
\usepackage{color}
\usepackage{amsthm} 

\allowdisplaybreaks

\newtheorem{definition}{Definition}

\newtheorem{theorem}[definition]{Theorem}

\newtheorem{lemma}[definition]{Lemma}

\def\QED{\mbox{\rule[0pt]{1.5ex}{1.5ex}}}

\def\endproof{\hspace*{\fill}~\QED\par\endtrivlist\unskip}

\newcommand{\Tr}{{\rm Tr}\,}

\makeatletter
\newcommand{\lleq}{\mathrel{\mathpalette\gl@align<}}
\newcommand{\ggeq}{\mathrel{\mathpalette\gl@align>}}
\newcommand{\gl@align}[2]{
\vbox{\baselineskip\z@skip\lineskip\z@
\ialign{$\m@th#1\hfil##\hfil$\crcr#2\crcr{}_{{}_{(=)}}\crcr}}}
\makeatother

\begin{document}
\title{Quantum Merlin-Arthur with Clifford Arthur} 
\author{Tomoyuki Morimae}
\email{morimae@gunma-u.ac.jp}
\affiliation{ASRLD Unit, Gunma University, 1-5-1 Tenjincho, Kiryushi,
Gunma, 376-0052, Japan}
\author{Masahito Hayashi}
\email{masahito@math.nagoya-u.ac.jp}
\affiliation{Graduate School of Mathematics, Nagoya University, Furocho,
Chikusaku, Nagoya, 464-8602, Japan}
\affiliation{Center for Quantum Technologies, National University of Singapore,
117543, Singapore}
\author{Harumichi Nishimura}
\email{hnishimura@math.cm.is.nagoya-u.ac.jp}
\affiliation{
Graduate School of Information Science, 
Nagoya University,
Furhocho, Chikusaku, Nagoya, Aichi, 464-8601 Japan}
\author{Keisuke Fujii}
\email{fujii.keisuke.2s@kyoto-u.ac.jp}
\affiliation{The Hakubi Center for Advanced Research, 
Kyoto University, YoshidaUshinomiyacho, Sakyoku, Kyoto 606-8302, Japan}
\affiliation{Graduate School of Science, Kyoto University, 
	Kitashirakawa, Oiwakecho, Sakyoku, Kyoto 606-8502, Japan}

\begin{abstract}
We show that the class QMA does not change even if
we restrict Arthur's computing ability to only Clifford gate operations
(plus classical XOR gate).
The idea is to use the fact that the preparation of certain
single-qubit states, so called magic states, plus any Clifford gate operations 
are universal for quantum computing. 
If Merlin is honest, he sends the witness plus magic states to Arthur.
If Merlin is malicious, he might send other states to Arthur, but
Arthur can verify the correctness of magic states by himself.
We also generalize the result to QIP[3]:
we show that
the class QIP[3] does not change even if
the computational power of the verifier is restricted to only 
Clifford gate operations (plus classical XOR gate).
\end{abstract}
\maketitle  

\section{Introduction}
In classical interactive proof, many results have been obtained
on the complexities of restricted verifiers. For example, Ref.~\cite{Con93}
surveys the studies of the case when the verifier is restricted to log-space computing.
In quantum interactive proof, on the other hand, we have more options for
restricting the verifier's ability. For example, we can
assume that the verifier can perform some restricted set of gates,
or even that the verifier is classical.
Most of the researches so far on such restricted quantum interactive proof
have been done for the multi-prover case or the case allowing
multi-communications between the prover and verifier~\cite{RUV,Ji15},
and therefore the simplest case, namely, a single prover and a single communication,
is not well understood.

The purpose of the present paper is to study the class QMA with a restricted verifier.
QMA (Quantum Merlin-Arthur) is a quantum analog of NP (or, more precisely,
MA (Merlin-Arthur)) defined by Kitaev~\cite{Kitaev}
and Watrous~\cite{Watrous} (also discussed by Knill~\cite{Knill}).
The prover, called Merlin, has unbounded
computational power and the verifier, called Arthur, can perform polynomial-time universal
quantum computing
by using a polynomial-size quantum state (so called a witness) 
sent from Merlin. For a yes instance, Arthur accepts the witness 
with high probability, and for a no instance, any Merlin's witness
is rejected by Arthur with high probability.
The formal definition of QMA is as follows:
\begin{definition}
A promise problem $A=(A_{yes},A_{no})$ is in QMA if and only if there exist polynomials
$p$, $q$, and a polynomial-time uniform family $\{Q_x\}$ of quantum circuits,
where $x\in A$ is the input with $|x|=n$, 
$Q_x$ takes as input a $p(n)$-qubit quantum state (so called the witness), 
and $q(n)$ ancilla qubits in state $|0\rangle^{\otimes q(n)}$,
such that
\begin{itemize}
\item[1.] Completeness: if $x\in A_{yes}$, then there exists a $p(n)$-qubit quantum
state $|w\rangle$ such that $Q_x$ accepts $|w\rangle$ with probability at
least $a$.
\item[2.]
Soundness: if $x\in A_{no}$, then for any $p(n)$-qubit quantum state $\xi$,
$Q_x$ accepts $\xi$ with probability at most $b$.
\end{itemize}
Here, $a-b\ge \frac{1}{poly(n)}$.
\end{definition}

In this definition, it is assumed that Arthur 
can perform universal quantum computing.
In this paper, we investigate what if Arthur is restricted to
apply only Clifford gate operations (plus universal classical computing).
Here, Clifford gate operations are operations generated by
$H\equiv|+\rangle\langle 0|+|-\rangle\langle1|$,
$S\equiv|0\rangle\langle 0|+i|1\rangle\langle1|$,
and
$CZ\equiv|0\rangle\langle 0|\otimes I+|1\rangle\langle1|\otimes Z$,
where $|\pm\rangle\equiv\frac{1}{\sqrt{2}}(|0\rangle\pm|1\rangle)$,
$I\equiv|0\rangle\langle0|+|1\rangle\langle1|$ is the two-dimensional identity operator,
and $Z\equiv|0\rangle\langle 0|-|1\rangle\langle1|$ is the Pauli $Z$ operator.
In this restriction, Arthur's computational power is restricted to be classical
in some sense~\cite{Nishimura_note}
as the Gottesman-Knill theorem~\cite{GK} says that
Clifford gate operations (plus universal classical computing) are classically simulable.
We show that such a restriction nevertheless does not change the power of QMA.
In other words, we show 

\begin{theorem}
${\rm QMA}_{\rm Clifford}={\rm QMA}$.
\end{theorem}
Here, ${\rm QMA}_{\rm Clifford}$ is defined as follows:

\begin{definition}
\label{def:QMA_Clifford}
The definition of ${\rm QMA}_{\rm Clifford}$ 
is the same as that of QMA except that
``a polynomial-time uniform family $\{Q_x\}$ of quantum circuits"
is replaced with
``a polynomial-time uniform family $\{V_x\}$ of quantum circuits that consist
of 
\begin{itemize}
\item[1.]
Preparation of $|0\rangle$.
\item[2.]
Measurements in the $Z$ basis (at any time during the computation).
\item[3.]
Clifford gates (that can be classically
controlled by the previous measurement results)."
\end{itemize}
Note that, for simplicity, we assume that Arthur can also perform
classical XOR gate. 
It is known that the generation of the three-qubit GHZ state (which
can be done with the preparation of $|0\rangle^{\otimes 3}$, 
and applications of $H$ and $CZ$), 
adaptive Pauli measurements (which can be done with the classically controlled
$H$, and the $Z$-basis measurements), and the classical XOR gate 
are universal for classical computing~\cite{Janet}.
\end{definition}

Our idea to show the theorem is to use the fact that the preparation
of many (i.e., polynomial in the input size) copies of the single-qubit state,
\begin{eqnarray*}
|H\rangle\langle H|\equiv\frac{1}{2}\Big[
I-\frac{1}{\sqrt{2}}(X+Z)\Big]
=\Big(\sin\frac{\pi}{8}|0\rangle-\cos\frac{\pi}{8}|1\rangle\Big)
\Big(\sin\frac{\pi}{8}\langle0|-\cos\frac{\pi}{8}\langle1|\Big),
\end{eqnarray*}
so called a magic state, plus any Clifford gate operations are universal
for quantum computing~\cite{magic,Fujii_note}.
Here, $X\equiv|0\rangle\langle1|+|1\rangle\langle0|$ is the Pauli $X$ operator.
Therefore, Arthur needs only Clifford gate operations if he asks
Merlin to add magic states to the witness.
One problem is that, for a no instance, Arthur cannot trust Merlin:
Merlin might send other states pretending to be sending magic states.
Therefore, Arthur has to do some test 
such that if the state sent from Merlin passes the test, 
then the output of the test is guaranteed to be close to magic states
with a sufficiently small significance level. 
We show that such a test does exist,
and therefore ${\rm QMA}_{\rm Clifford}={\rm QMA}$.

The idea can also be applied to QIP (Quantum Interactive Proof), 
which is a generalization of QMA
where many quantum messages can be exchanged between Merlin (the prover) 
and Arthur (the verifier).
QIP was defined by Watrous in Ref.~\cite{Wat99}, and it is known that
${\rm QIP}[3]={\rm QIP}={\rm PSPACE}$~\cite{KW00,JJUW09},
where QIP[$k$] means that
the prover and the verifier can exchange quantum messages $k$ times
(hence ${\rm QMA}={\rm QIP[1]}$),
and ${\rm QIP}\equiv\cup_{k=poly(n)} {\rm QIP}[k]$.
We show ${\rm QIP}[3]_{\rm Clifford}={\rm QIP[3]}$,
where ${\rm QIP}[3]_{\rm Clifford}$ is defined in a similar way
as ${\rm QMA}_{\rm Clifford}$:
the verifier of ${\rm QIP[3]}$ is restricted to only Clifford gate operations
(plus classical XOR gate).

Finally, it is interesting to compare our result on QMA
with QCMA. 
QCMA is a variant of QMA where the witness sent from Merlin
is not a quantum state but a classical bit string.
Since the three operations in Definition~\ref{def:QMA_Clifford}
are classically simulable (the Gottesman-Knill theorem~\cite{GK}),
we obtain ${\rm QCMA}_{\rm Clifford}\subseteq{\rm MA}$.
Here, ${\rm QCMA}_{\rm Clifford}$ is defined in a similar way as ${\rm QMA}_{\rm Clifford}$:
Arthur is restricted to Clifford gates (plus the classical XOR gate).

\section{Magic state test}

Let us consider the following test,
which we call {\it the magic state test}:
\begin{itemize}
\item[1.]
Let $\Omega_1$ be a $(2r(n)+s(n)+l(n)+p(n))$-qubit system, where 
$r$, $s$, $l$, and $p$ are polynomials specified later.
\item[2.]
Let $\Omega_2$ be the subsystem of $\Omega_1$ consisting
of the first $(2r(n)+s(n)+l(n))$ qubits of $\Omega_1$.
\item[3.]
We randomly choose $(2r(n)+s(n))$ qubits from $\Omega_2$.  
Let $\Omega_3$ be the system of thus chosen $(2r(n)+s(n))$ qubits.
\item[4.]
We further randomly choose $2r(n)$ qubits from $\Omega_3$,
and divide thus chosen $2r(n)$ qubits into two $r(n)$-qubit groups, $S_1$ and $S_2$.
We measure each qubit of $S_1$ ($S_2$, resp.) in
$X$ ($Z$, resp.) basis. 
Let $x$ and $z$ be the number of obtaining $+1$ results for
$X$ and $Z$ measurements, respectively.
If $x$ and $z$ are larger than $F(\delta_1,\delta_2,r(n))$, the test is passed,
where $F(\delta_1,\delta_2,r(n))$ is the maximum value satisfying
\begin{eqnarray*}
\delta_1\ge \sum_{k=0}^{F(\delta_1,\delta_2,r(n))}{r(n)\choose k}
\Big(\frac{1}{2}-\frac{1}{2\sqrt{2}}+\delta_2\Big)^k
\Big(\frac{1}{2}+\frac{1}{2\sqrt{2}}-\delta_2\Big)^{r(n)-k}.
\end{eqnarray*}
Here, $\delta_1$ and $\delta_2$ are specified later.
\item[5.]
Let $\sigma$ be the state of $s(n)$ qubits of $\Omega_3$ that were not measured.
\end{itemize}

We can show that the correct magic states pass the
magic state test with high probability,
and that if we pass the magic state test,
$\sigma$ is close to the correct magic states.
More precisely, we can show the following lemma. (Its proof is given in
Appendix~\ref{app1}.)
\begin{lemma}
\label{Hayashi_lemma}
We take 
$
\delta_2=\frac{2\delta_1}{\sqrt{2}s(n)}, 
$
and choose $\delta_1$ as
$
\delta_1 \le\frac{1}{4000}.
$
We also take $r(n)$ as
\begin{eqnarray}
r(n)=
\Big(\frac{\sqrt{2}s(n)}{2\delta_1}
\sqrt{8} (\Phi^{-1}(\epsilon)+\Phi^{-1}(\delta_1))\Big)^2, 
\label{BB}
\end{eqnarray}
where $\epsilon$ is any constant, 
and
\begin{eqnarray*}
\Phi(x)\equiv\int_{x}^{\infty}\frac{1}{\sqrt{2\pi}}e^{-\frac{t^2}{2}}dt.
\end{eqnarray*}
Finally, we take $l$ as
\begin{eqnarray}
\sqrt{\frac{2(2r(n)+s(n)-1)^2\log2}{l(n)}}\le\frac{1}{2000}.
\label{CC}
\end{eqnarray}
Then, we have the following items.
\begin{description}
\item[(i)]
If 
\begin{eqnarray*}
\rho=|H\rangle\langle H|^{\otimes 2r(n)+s(n)+l(n)}\otimes \xi,
\end{eqnarray*}
we pass the magic state test with probability $1-\epsilon-o(1)$ for any $p(n)$-qubit state $\xi$.
\item[(ii)]
Furthermore, for any state $\rho$ of $\Omega_1$, if we pass the magic state test,
we can guarantee that
\begin{eqnarray}
\langle H^{\otimes s(n)}|\sigma|H^{\otimes s(n)}\rangle
\ge 1-\frac{1}{100} \label{ab}
\end{eqnarray}
with the significance level $\frac{1}{10}$.
\end{description}
\end{lemma}
(Note that the significance level is
the maximum passing probability 
when malicious Merlin sends incorrect states 
so that the resultant state $\sigma$ does not satisfy Eq.~(\ref{ab})~\cite{textbook}.
)

We can also show the following lemma
(its proof is given in Appendix~\ref{app2}),
which will be used later:
\begin{lemma}
\label{Hayashi_lemma2}
Let $\rho$ be a state in ${\mathcal H}_1\otimes {\mathcal H}_2$, 
where ${\mathcal H}_1$ and ${\mathcal H}_2$ are Hilbert spaces.
For any $|x\rangle\in {\mathcal H}_1$,
\begin{eqnarray}
\max_{\rho'\in {\mathcal H}_2}F(|x\rangle\langle x|\otimes \rho',\rho)^2=F(|x\rangle\langle x|,
{\rm Tr}_2(\rho))^2,
\label{l2}
\end{eqnarray}
where $F(\sigma_1,\sigma_2)\equiv{\rm Tr}|\sqrt{\sigma_1}\sqrt{\sigma_2}|$
is the fidelity between $\sigma_1$ and $\sigma_2$,
and ${\rm Tr}_2$ is the partial trace over ${\mathcal H}_2$.
\end{lemma}

\section{Proof of the Theorem}
Now let us show our main result.

{\it Proof of Theorem 2}:
${\rm QMA}_{\rm Clifford}\subseteq{\rm QMA}$
is obvious. Let us show
${\rm QMA}_{\rm Clifford}\supseteq{\rm QMA}$.
We assume that a promise problem $A$ is in 
${\rm QMA}$.
Then, there exist a polynomial-time uniform family $\{Q_x\}$ of quantum circuits,
and the $p(n)$-qubit witness state $|w\rangle$ that is accepted by Arthur with 
probability at least $a$ if $x\in A_{yes}$,
while any $p(n)$-qubit state is accepted with probability at most $b$
if $x\in A_{no}$. Here we can take $a=\frac{2}{3}$ and $b=\frac{1}{3}$.
Let $V_x$ be a quantum circuit satisfying the conditions of Definition~\ref{def:QMA_Clifford}
and simulating $Q_x$ exactly by the method in Ref.~\cite{magic}, and let $s(n)$
be the number of magic states consumed for this simulation.

Arthur runs the following protocol:
\begin{itemize}
\item[1.]
Arthur receives a $(2r(n)+s(n)+l(n)+p(n))$-qubit state $\rho$ from Merlin.
If Merlin is honest, 
\begin{eqnarray*}
\rho=|H\rangle\langle H|^{\otimes 2r(n)+s(n)+l(n)}\otimes|w\rangle\langle w|.
\end{eqnarray*}
If he is malicious, $\rho$ can be any state.
\item[2.]
Arthur does the magic state test on $\rho$.
\item[3.]
If $\rho$ fails to pass the test, Arthur rejects.
\item[4.]
If $\rho$ passes the test, Arthur now has an $(s(n)+p(n))$-qubit state.
The first $s(n)$ qubits are used as magic states to simulate
$Q_x$ with $V_x$, 
and the state of the last $p(n)$ qubits 
is used as the witness for $Q_x$.
\end{itemize}

First, we consider the case when
$x\in A_{yes}$. 
In this case, Merlin sends correct magic states, and therefore
the probability of passing the test
is $1-\frac{1}{10}$ from Lemma~\ref{Hayashi_lemma},
where we take $\epsilon=\frac{1}{10}$.
Therefore, Arthur's acceptance
probability $p_{acc}$ is 
\begin{eqnarray*}
p_{acc}\ge a\times \Big(1-\frac{1}{10}\Big)=\frac{9a}{10}\equiv a'.
\end{eqnarray*}

Next let us consider the case when
$x\in A_{no}$. 
Arthur's acceptance probability $p_{acc}$ is  
\begin{eqnarray*}
p_{acc}=\mbox{Tr}(C_x\eta)\times P(\mbox{pass the test})
\end{eqnarray*}
for a certain POVM element $C_x$ such that the corresponding POVM depends on $x$
and is implementable with only Clifford gates,
where $\eta$ is the $(s(n)+p(n))$-qubit state after passing the magic state test,
and $P(\mbox{pass the test})$ is the probability of passing the 
magic state test.
From Eqs.~(\ref{ab}) and (\ref{l2}), and the relation,
\begin{eqnarray*}
	\frac{1}{2}\|\rho_1-\rho_2\|_1
	\le\sqrt{1-F(\rho_1,\rho_2)^2},
\end{eqnarray*}
between the fidelity and the trace distance 
(e.g., Eq.~(6.106) of Ref.~\cite{Hayashi_book}),
$\eta$ satisfies
\begin{eqnarray*}
\frac{1}{2}\Big\|\eta-|H\rangle\langle H|^{\otimes s(n)}\otimes\xi\Big\|_1
\le \frac{1}{10}
\end{eqnarray*}
with probability $1-\frac{1}{10}$
for a certain $p(n)$-qubit state $\xi$.
Then,
\begin{eqnarray*}
\mbox{Tr}(C_x\eta)
-\mbox{Tr}\Big[C_x(|H\rangle\langle H|^{\otimes s(n)}\otimes \xi)\Big]
&\le&
\Big|\mbox{Tr}(C_x\eta)
-\mbox{Tr}\Big[C_x(|H\rangle\langle H|^{\otimes s(n)}\otimes \xi)\Big]\Big|\\
&\le&
\frac{1}{2}\Big\|\eta-|H\rangle\langle H|^{\otimes s(n)}\otimes \xi\Big\|_1\\
&\le&\frac{1}{10}.
\end{eqnarray*}
Therefore,
\begin{eqnarray*}
p_{acc}&=&
\mbox{Tr}(C_x\eta)P(\mbox{pass the test})\\
&\le&
\mbox{Tr}(C_x\eta)\\
&\le&\frac{9}{10}\Big(\mbox{Tr}\Big[
C_x(|H\rangle\langle H|^{\otimes s(n)}\otimes\xi)\Big]+\frac{1}{10}\Big)
+\frac{1}{10}\\
&\le&\frac{9b}{10}+\frac{9}{100}+\frac{1}{10}
\equiv b'.
\end{eqnarray*}

Since $a=\frac{2}{3}$ and $b=\frac{1}{3}$,
\begin{eqnarray*}
a'-b'&=&\frac{9a}{10}-\frac{9b}{10}-\frac{9}{100}-\frac{1}{10}\\
					&=&\frac{11}{100}.
\end{eqnarray*}
Therefore,
$L$ is in ${\rm QMA}_{\rm Clifford}$.

\section{QIP}
We can apply our idea to QIP.
Let us consider ${\rm QIP}[3]$. 
First, the prover applies a unitary map ${\mathcal P_1}$ on
the $\alpha(n)$-qubit state $|0\rangle_P^{\otimes \alpha(n)}$, so
called the prover's private register, and the $\beta(n)$-qubit state
$|0\rangle_M^{\otimes \beta(n)}$, so called the message register,
where $\alpha$ and $\beta$ are some polynomials.
The prover sends the message register of 
\begin{eqnarray*}
{\mathcal P}_1\Big(\fbox{0}_P^{\otimes\alpha(n)}\otimes\fbox{0}_M^{\otimes\beta(n)}\Big)
\end{eqnarray*}
to the verifier, where
we have used the notation $\fbox{$x$}\equiv|x\rangle\langle x|$.
The verifier applies a unitary map ${\mathcal V_1}$ on the message register
plus the $\gamma(n)$-qubit state $|0\rangle_V^{\otimes \gamma(n)}$,
so called the verifier's private register:
\begin{eqnarray*}
{\mathcal V}_1\Big({\mathcal P}_1
\big(\fbox{0}_P^{\otimes \alpha(n)}\otimes\fbox{0}_M^{\otimes\beta(n)}\big)
\otimes\fbox{0}_V^{\otimes\gamma(n)}\Big),
\end{eqnarray*}
where $\gamma$ is a polynomial.
The verifier sends the message register to the 
prover, and the prover returns it
after applying a unitary map ${\mathcal P}_2$ on the message register plus 
the prover's private register.
Now they share the state
\begin{eqnarray*}
{\mathcal P}_2{\mathcal V}_1\Big({\mathcal P}_1
\big(\fbox{0}_P^{\otimes\alpha(n)}\otimes\fbox{0}_M^{\otimes\beta(n)}\big)\otimes
\fbox{0}_V^{\otimes \gamma(n)}\Big),
\end{eqnarray*}
where the message register is possessed by the verifier.
Finally, the verifier performs a POVM measurement 
on the verifier's private register plus the message register
in order to decide the acceptance or
rejection.

In the case of ${\rm QIP}[3]_{\rm Clifford}$,
the verifier performs the magic state test, 
which we denote ${\mathcal T}$, on
(the message part of) the state 
${\mathcal P}_1\big(\fbox{0}_P^{\otimes \alpha(n)}\otimes\fbox{0}_M^{\otimes \beta(n)}\big)$.
Let $C_x$ be the POVM element applied by the verifier
that corresponds to the acceptance.
(The corresponding POVM depends on $x$ and is implementable with only
Clifford gate operations.)
Then, in the similar way as in the case of QMA, we can show
\begin{eqnarray*}
&&\mbox{Tr}\Big[C_x {\mathcal P}_2{\mathcal V}_1\Big(
{\mathcal T}{\mathcal P}_1\big(\fbox{0}_P^{\otimes \alpha}\otimes\fbox{0}_M^{\otimes \beta}\big)
\otimes\fbox{0}_V^{\otimes \gamma}
\Big)\Big]
-
\mbox{Tr}\Big[C_x {\mathcal P}_2{\mathcal V}_1\Big(
\big(\fbox{$H$}^{\otimes s}\otimes \xi\big)\otimes\fbox{0}_V^{\otimes \gamma}
\Big)\Big]\\
&\le&
\frac{1}{2}\Big\|
{\mathcal P}_2{\mathcal V}_1\Big({\mathcal T}
{\mathcal P}_1\big(\fbox{0}_P^{\otimes \alpha}\otimes\fbox{0}_M^{\otimes \beta}\big)
\otimes\fbox{0}_V^{\otimes \gamma}
\Big)
-
{\mathcal P}_2{\mathcal V}_1\Big(
\big(\fbox{$H$}^{\otimes s}\otimes \xi\big)\otimes\fbox{0}_V^{\otimes \gamma}
\Big)\Big\|_1\\
&=&
\frac{1}{2}\Big\|
{\mathcal T}{\mathcal P}_1
\big(\fbox{0}_P^{\otimes \alpha}\otimes\fbox{0}_M^{\otimes \beta}\big)
-
\fbox{$H$}^{\otimes s}\otimes \xi
\Big\|_1\\
&\le&\frac{1}{10},
\end{eqnarray*}
where $\xi$ is a state of a part of the message register and
prover's private register.
Then by using a similar argument of QMA, we can show 
${\rm QIP}[3]\subseteq{\rm QIP}[3]_{\rm Clifford}$.

For ${\rm QIP}[2]$, we do not know whether
${\rm QIP}[2]={\rm QIP}[2]_{\rm Clifford}$ holds,
since in this case, the verifier has to perform a unitary map first,
and no magic state is available for the first unitary map.

\acknowledgements
TM is supported by the JSPS Grant-in-Aid for Young Scientists (B) No.26730003 and 
the MEXT JSPS Grant-in-Aid for Scientific Research on Innovative Areas No.15H00850.
MH is partially supported by the 
JSPS Grant-in-Aid for Scientific Research (A) No. 23246071 and the National
Institute of Information and Communication Technology
(NICT), Japan. The Centre for Quantum Technologies is
funded by the Singapore Ministry of Education and the National
Research Foundation as part of the Research Centres of Excellence programme.
HN is supported by the JSPS Grant-in-Aid for Scientific Research (A) 
Nos.23246071, 24240001, 26247016,
and (C) No.25330012,
and the MEXT JSPS Grant-in-Aid for Scientific Research on
Innovative Areas No.24106009.

\appendix

\section{Proof of Lemma~\ref{Hayashi_lemma}}
\label{app1}

\noindent{\it Proof of (i):}\quad
The condition \eqref{BB} is equivalent with
\begin{align}
\frac{\sqrt{2}s}{2\delta_1}
\sqrt{8} (\Phi^{-1}(\epsilon)+\Phi^{-1}(\delta_1))
=
\sqrt{r}, \label{g7}
\end{align}
i.e.,
\begin{align*}
\sqrt{8r} (\Phi^{-1}(\epsilon)+\Phi^{-1}(\delta_1))
=
r\frac{2\delta_1}{\sqrt{2}s}.
\end{align*}
This condition implies that
\begin{eqnarray*}
r \Big(\frac{1}{2}-\frac{1}{2\sqrt{2}}\Big)
+\sqrt{8r} \Phi^{-1}(\epsilon)
&=&
r \Big(\frac{1}{2}-\frac{1}{2\sqrt{2}}\Big)
-\sqrt{8r} \Phi^{-1}(1-\epsilon)\\
&=&
r \Big(\frac{1}{2}-\frac{1}{2\sqrt{2}}\Big)
+r\frac{2\delta_1}{\sqrt{2}s}
-\sqrt{8r} \Phi^{-1}(\delta_1) 
\end{eqnarray*}
Due to the central limit theorem, 
the RHS asymptotically equals
the quantity $F(\delta_1,\delta_2,r)$ as
\begin{align*}
F(\delta_1,\delta_2,r)
&\cong
r \Big(\frac{1}{2}-\frac{1}{2\sqrt{2}}+\delta_2\Big)
-\sqrt{\frac{r}{
(\frac{1}{2}-\frac{1}{2\sqrt{2}}+\delta_2)
(\frac{1}{2}+\frac{1}{2\sqrt{2}}-\delta_2)
}
} \Phi^{-1}(\delta_1)
\\
&\cong
r \Big(\frac{1}{2}-\frac{1}{2\sqrt{2}}+\frac{2\delta_1}{\sqrt{2}s}\Big)
-\sqrt{\frac{r}{
(\frac{1}{2}-\frac{1}{2\sqrt{2}})
(\frac{1}{2}+\frac{1}{2\sqrt{2}})
}
} \Phi^{-1}(\delta_1)
\\
&=r \Big(\frac{1}{2}-\frac{1}{2\sqrt{2}}\Big)
+r\frac{2\delta_1}{\sqrt{2}s}
-\sqrt{8r} \Phi^{-1}(\delta_1),
\end{align*}
where $\cong$ means that both sides equal up to a $O(1)$ additive factor.
Hence, 
again due to the central limit theorem,
the accepting probability with the true state 
$|H\rangle^{\otimes (2r+s+l)}  $ is calculated as
\begin{align*}
&\sum_{x=0}^{F(\delta_1,\delta_2,r)}
{r \choose x}
\Big(\frac{1}{2}-\frac{1}{2\sqrt{2}}\Big)^x
\Big(\frac{1}{2}+\frac{1}{2\sqrt{2}}\Big)^{r-x} \\
 \cong &
\sum_{x=0}^{r \Big(\frac{1}{2}-\frac{1}{2\sqrt{2}}\Big)
+\sqrt{8r} \Phi^{-1}(\epsilon)}
{r \choose x}
\Big(\frac{1}{2}-\frac{1}{2\sqrt{2}}\Big)^x
\Big(\frac{1}{2}+\frac{1}{2\sqrt{2}}\Big)^{r-x} 
 \cong 
1-\epsilon ,
\end{align*}
which implies Item (i),
where $\cong$ means that both sides equal up to a $o(1)$ additive factor.

\noindent{\it Proof of (ii):}\quad
To show Item (ii),
let $P'$ be the POVM element on 
the composite system of $S_1$ and $S_2$ corresponding to passing the test.
We define the operator 
$P\equiv P' \otimes (I- | H^{\otimes s}\rangle \langle H^{\otimes s}|)$
that corresponds to the incorrect decision.
To bound the probability of incorrect decision,
we prepare the following lemma.
\begin{lemma}
\label{app_lemma}
Any state $\rho$ on the system $\mathbb{C}^{2r+s}$ satisfies 
\begin{align}
\Tr P \rho \le 
\max\{2\delta_1, s\sqrt{2}\delta_2\} +\sqrt{\frac{2(2r+s-1)^2 \log 2}{l}}.
\label{g3}
\end{align}
\end{lemma}
The proof of this lemma is given later.

The meaning of the value $\Tr P \rho $ is the following.
We fix a small real number $\delta>0$.
If the resultant state $\sigma'$ with the acceptance satisfies 
$\langle H^{\otimes s}|  \sigma' | H^{\otimes s}\rangle \le 1-\delta$,
i.e., 
$\Tr \sigma'  (I- | H^{\otimes s}\rangle \langle H^{\otimes s}| )
\ge \delta$,
the probability of the acceptance is less than 
$\frac{1}{\delta} \Tr P \rho $
because 
$\Tr P \rho \ge \delta \Tr (P' \otimes I )\rho $.
In other words, 
the resultant state $\sigma'$ with the acceptance satisfies 
$\langle H^{\otimes s}|  \sigma' | H^{\otimes s}\rangle > 1-\delta$
with the significance level 
$\max_{\rho}\frac{1}{\delta} \Tr P \rho $.

Let us choose $\delta_2$ as $\frac{2\delta_1}{\sqrt{2}s}$, 
and $\delta$ as $\frac{1}{100}$.
Then,
\begin{align}
\Tr P \rho \le 
2\delta_1 +\sqrt{\frac{2(2r+s-1)^2 \log 2}{l}}.
\label{g5}
\end{align}
Combining the relation \eqref{CC} and $\delta_1\leq\frac{1}{4000}$, 
we can evaluate the significance level as
$\max_{\rho}\frac{1}{\delta} \Tr P \rho \le \frac{1}{10}$, i.e., 
we obtain Item (ii).

\noindent{\it Proof of Lemma~\ref{app_lemma}:}
Firstly, we consider the case when 
the true state $\rho$ is a tensor product state $\sigma^{\otimes 2r+s}$.
When $x$ and $z$ are larger than 
$F(\delta_1,\delta_2,r)$,
we can conclude that 
\begin{align}
	\Tr (\sigma |+\rangle \langle +|)
&\le \frac{1}{2}-\frac{1}{2\sqrt{2}}+\delta_2 \label{a},\\
	\Tr (\sigma |0\rangle \langle 0|)
&\le \frac{1}{2}-\frac{1}{2\sqrt{2}}+\delta_2 \label{b},
\end{align}
with the significance level $2 \delta_1$.
So, if \eqref{a} and \eqref{b} do not hold,
\begin{align}
\Tr \sigma^{\otimes 2r} P' \le 2 \delta_1. \label{e}
\end{align}
On the other hand, 
the relations~\eqref{a} and \eqref{b} are equivalent with 
\begin{align}
	\Tr (\sigma (|+\rangle \langle +|-|-\rangle \langle -|))
&\le -\frac{1}{\sqrt{2}}+2\delta_2 \label{a2},\\
	\Tr (\sigma (|0\rangle \langle 0|-|1\rangle \langle 1|))
&\le -\frac{1}{\sqrt{2}}+2\delta_2 \label{b2}.
\end{align}
That is,
\begin{align}
\Tr \sigma | H\rangle\langle H|
=
\Tr \sigma 
\frac{1}{2}\Big[
I-\frac{1}{\sqrt{2}}(X+Z)\Big]
\ge 
1 - \sqrt{2}\delta_2,
\end{align}
which implies that
\begin{align}
\langle H^{\otimes s}|\sigma^{\otimes s} | H^{\otimes s}\rangle
\ge 
(1 - \sqrt{2}\delta_2)^s
\ge
1 - s\sqrt{2}\delta_2.\label{c}
\end{align}
Therefore, if \eqref{a} and \eqref{b} hold,
\eqref{c} holds, i.e.,
\begin{align}
\Tr \sigma^{\otimes s} 
(I- | H^{\otimes s}\rangle \langle H^{\otimes s}|)
\le s\sqrt{2}\delta_2.\label{f}
\end{align}
Combining \eqref{e} and \eqref{f} implies that
\begin{align}
\Tr P \sigma^{\otimes 2r+s}  
=
(\Tr \sigma^{\otimes 2r} P')\cdot
(\Tr \sigma^{\otimes s} 
(I- | H^{\otimes s}\rangle \langle H^{\otimes s}|))
\le 
\max\{2\delta_1, s\sqrt{2}\delta_2\} .\label{g1}
\end{align}

Since we randomly choose samples,
we can assume that the total system is permutation invariant.
Hence, we can apply the quantum de Finetti theorem to $\Omega_3$.
In particular, our measurements are one-way LOCC.
So, we can apply the equation (2) in \cite{LS}.
For the state $\rho$ in $\Omega_3$,
there exists a distribution $Q$ on the qubit space ${\cal S}(\mathbb{C}^2) $
such that
\begin{align}
\Big|\Tr P \rho -\Tr P \int Q(\sigma) \sigma^{\otimes 2r+s} d\sigma \Big|
\le
\sqrt{\frac{2(2r+s-1)^2 \log 2}{l}}.\label{g2}
\end{align}
\eqref{g1} yields that
\begin{align}
\Tr P \int Q(\sigma) \sigma^{\otimes 2r+s} d\sigma
\le 
\max\{2\delta_1, s\sqrt{2}\delta_2\} .\label{g4}
\end{align}
Thus, \eqref{g2} and \eqref{g4} guarantee \eqref{g3}.
\endproof

\section{Proof of Lemma~\ref{Hayashi_lemma2}}
\label{app2}

Note that $\langle x|{\rho}|x\rangle$ is a non-negative hermitian matrix on ${\mathcal H}_2$.
Let us define the state 
\begin{align*}
\rho''\equiv
\frac{1}{\Tr_2 \langle x|{\rho}|x\rangle}
\langle x|{\rho}|x\rangle
=\frac{1}{ \langle x|\Tr_2{\rho}|x\rangle}
\langle x|{\rho}|x\rangle.
\end{align*}
Then,
\begin{eqnarray*}
F(|x\rangle \langle x| \otimes \rho',\rho)^2
&=&\Big(\Tr| \sqrt{|x\rangle \langle x| \otimes \rho'} \sqrt{\rho}|\Big)^2\\
&=&\Big(\Tr| |x\rangle \langle x| \otimes \sqrt{\rho'} \sqrt{\rho}|\Big)^2\\
&=&\Big(\Tr\sqrt{ 
|x\rangle \langle x| \otimes \sqrt{\rho'} {\rho}
|x\rangle \langle x| \otimes \sqrt{\rho'} 
}\Big)^2 \\
&=&\Big(\Tr|x\rangle\langle x|  \otimes
\sqrt{ 
\sqrt{\rho'} \langle x|{\rho}|x\rangle \sqrt{\rho'} }
\Big)^2 \\
&=&\Big(\Tr \sqrt{\sqrt{\rho'} \langle x|{\rho}|x\rangle \sqrt{\rho'} }
\Big)^2 \\
&=&\Big(\Tr \sqrt{\sqrt{\rho'} \rho''\sqrt{\rho'}  \langle x|\Tr_2{\rho}|x\rangle }
\Big)^2 \\
&=&\Big(\Tr \sqrt{\sqrt{\rho'} \rho''\sqrt{\rho'} }\Big)^2
 \langle x|\Tr_2{\rho}|x\rangle 
 \\
	&=&F(\rho', \rho'')^2
 \langle x|\Tr_2{\rho}|x\rangle \\
	&\le& \langle x|\Tr_2{\rho}|x\rangle,
\end{eqnarray*}
where the equality holds when 
$\rho'= \rho''$.


\begin{thebibliography}{00}
\bibitem{Nishimura_note}
Arthur's power is also regarded to be quantum in that he can do Clifford operations
for quantum superposition, while if he cannot deal with quantum superposition (except
the measurement in the computational basis), the power of QMA with such Arthur trivially
collapses to MA.
\bibitem{Fujii_note}
Note that the preparation of $|0\rangle$, $S$, CNOT,
measurements of $X$ and $Z$, and the preparation of $|H\rangle$
are also universal. Furthermore, the CNOT can be mutually commuting,
which means that we divide the register into two subsets, and 
qubits in the one set are always controlled ones,
and those in the other set are always target ones.
The proof of the universality is the same as that of the topological
quantum computing on the surface code, where we can implement only
Abelian anyons, i.e., commuting CNOT.
\end{thebibliography}
\end{document}